\documentclass[12pt]{article}
\usepackage{sc3conf}
\usepackage{amsfonts}

\def\ads{AdS_{5}}
\def\be{\begin{equation}}
\def\ee{\end{equation}}
\def\ba{\begin{array}}
\def\ea{\end{array}}
\def\d{\partial}
\def\dps{\displaystyle}

\def\ba{\begin{array}}
\def\ea{\end{array}}
\def\d{\partial}
\def\dps{\displaystyle}

\def\bpsi{\bar{\psi}}

\begin{document}
\begin{center}
{\it Presented at the Third International Sakharov Conference on Physics,\\
Moscow, June 2002.}
\end{center}
\vspace{1cm}
\title{Cubic SUSY Interactions of Higher Spin Gauge Fields in $AdS_5$ }

\authors{K.B. Alkalaev}

\addresses{I.E.Tamm Department of Theoretical Physics, Lebedev Physical Institute,\\
Leninsky prospect 53, 119991, Moscow, Russia}

\maketitle

\begin{abstract}

This contribution discusses  the recent progress in research of consistent
supersymmetric interactions of $AdS_5$ higher spin gauge fields.
\end{abstract}

\vspace{1cm}

The issue of consistent interactions between higher spin gauge
fields is certainly not new one. Originated as the quest for an extension
of ordinary supergravities, higher spin gauge theories
are mainly motivated now by their possible relationship with a symmetric phase of a
theory of fundamental interactions presently identified with M theory.
However,
in spite of the considerable progress in constructing higher
spin gauge theories (for review see \cite{V_obz2}) we are still
missing some the basic ingredients. In particular, a non-trivial
problem which remains unsolved is how to write down an action
principle beyond the cubic order in interactions.
Nonetheless, some essential features of any higher
spin gauge theory such as an algebra of global higher spin symmetries
and the crucial relevance of $AdS$ background may be revealed
already within cubic theory presently available in $d\leq 5$.

In this contribution we review general construction of cubic
supersymmetric interactions of totally symmetric bosonic and
fermionic higher spin gauge fields propagating on $AdS_5$
background \cite{Alkalaev:2002rq}. The theory incorporates the
previous results for purely bosonic interactions \cite{VD5} and
naturally extends to five dimensions the method for constructing
cubic higher spin action originally developed in $d=4$ \cite{FV1}.

We begin with identifying the algebra of $AdS_5$ global higher spin symmetries.
To this end, consider the associative Weyl-Clifford algebra with
non-vanishing (anti)commutation generating relations ($\alpha,\beta=1\div4$)
\be
\label{spa}
[a_{\alpha}, b^{\beta}]_\star=\delta_{\alpha}{}^{\beta}\;,
\qquad \{\psi, \bpsi\}_\star=1\;,
\ee
with respect to Weyl star product. The generators
\be
\label{gl}
T_\alpha{}^\beta = a_\alpha b^\beta \;,\qquad
Q_\alpha=a_\alpha\bpsi\;, \qquad \bar{Q}^\beta= b^\beta\psi\;,
\qquad U=\psi\bpsi\;
\ee
close to the superalgebra $gl(4|1;{\bf C})$ with respect
to the graded Lie supercommutator. The set of generators (\ref{gl}) consists of $gl(4;{\bf C})$ generators
$T$, supersymmetry generators $Q$ and $\bar{Q}$ and $u(1)$ generator $U$.
The central element in $gl(4|1;{\bf C})$ is
$N = a_\alpha b^\alpha -\psi\bpsi$. The supertraceless part of (\ref{gl})
defines the generators of $sl(4|1;{\bf C})$, while the $\ads$ superalgebra $su(2,2|1)$
is a real form of $sl(4|1;{\bf C})$ singled out by the appropriate reality
conditions \cite{VD5}.

A natural higher spin extension of $su(2,2|1)$ introduced in
\cite{FLA} under the name $shsc^{\infty}(4|1)$ and called
$cu(1,1|8)$ in \cite{Vc}
is associated with the star product algebra of all polynomials
$F(a,b,\psi,\bpsi)$ satisfying the condition
\be
\label{dpoc}[N,F]_\star=0\;.
\ee
In other words, the $5d$ higher spin superalgebra $cu(1,1|8)$ is
spanned by star-(anti)com\-mutators of the elements of the centralizer
of $N$ in the star product algebra (\ref{spa}). Note that
$F\in cu(1,1|8)$ is  even in superoscillators.

The higher spin strength $ R(a, b, \psi, \bpsi |x)$ associated
with the higher spin connection $ \Omega(a, b, \psi, \bpsi |x)$ is
\be
\label{Rosn}
R= d\Omega + \Omega  \wedge\star\,
\Omega\,,\qquad d=dx^{\underline{n}}\frac{\d}{\d
x^{\underline{n}}}.
\ee
Infinitesimal higher spin gauge transformations are
\be
\label{gotr}
\delta\Omega =  D\epsilon\,,
\qquad \delta R =  [R  \,, \epsilon ]_\star\,,
\ee
where 0-form $\epsilon=\epsilon(a,b,\psi,\bar{\psi}|x)$ is an arbitrary
infinitesimal higher spin gauge symmetry parameter and
$\label{Dc} D F = d F + [\Omega  \,, F ]_\star.$
To analyse interactions we use the perturbation expansion
with the dynamical fields $\Omega_1$ treated as fluctuations above
the appropriately chosen background $\Omega_0$:
$\label{go01} \Omega = \Omega_0 +\Omega_1$,
where the vacuum gauge fields
$\Omega_0=\Omega_{0\,\beta}^{\;\;\alpha}(x)\, a_\alpha b^\beta$
correspond to background $AdS_5$ geometry described by virtue of
the zero-curvature condition $R(\Omega_0 )\equiv d\Omega_0+
\Omega_0 \wedge\star\, \Omega_0=0$
(for more details see \cite{VD5,Alkalaev:2002rq}).

Under the gauging procedure the algebra $cu(1,1|8)$ yields the set
of Lorentz higher spin fields which have different dynamical interpretation. In
the gauge sector one distinguishes between physical, auxiliary and
"extra" type fields. The auxiliary and "extra" fields carry the
ghost-type degrees of freedom and, by virtue of appropriately chosen
constraints \cite{lv,ss,VD5,Alkalaev:2002rq}, can be expressed in
terms of the physical fields  modulo pure gauge
degrees of freedom.
These constraints combined with the free higher spin equations of motion represent
the First On-Mass-Shell Theorem which plays a key role in constructing
consistent cubic interactions (see below).

The physical fields are arranged into an infinite sequence of
supermultiplets of totally symmetric higher spin gauge fields
$\{s\}^{(k)}, 0\leq k <\infty$, with a spin
content $(s,s-\frac{1}{2},s-1)^{(k)}$ determined by an integer
highest spin $s=2,3,...,\infty \;$.
The infinite degeneracy associated with
parameter $k$ is inherited from the fact that the corresponding higher spin
connection $ \Omega(a, b, \psi, \bpsi |x)$ is not supertraceless
and decomposes into an infinite series of supertraceless parts.
Algebraically, the origin of this infinite degeneracy  can be traced back to the fact
that the algebra $cu(1,1|8)$ is
not simple but contains infinitely many ideals generated by
the central element $N$.
One may consider quotient algebras and, in particular,
one of the most interesting reduction provided
by the algebra $hu_0(1,1|8)=cu(1,1|8)\!/I_N$, where $I_N$ is the
ideal spanned by the elements $N\star F = F\star N$ \cite{Vc,Alkalaev:2002rq}.

Strictly speaking, the theory
we consider is not fully supersymmetric because we truncate away
all lower spin fields with $s\leq 1$ (in particular, the spin 1
field from the spin 2 supermultiplet). This truncation is done to
simplify analysis because lower spin fields require special
formulation while our goal is to check consistency of the
higher-spin-gravitational interactions. By analogy with the $4d$
analysis (see second reference in  \cite{FV1}) it is not expected
to be a hard problem to extend our analysis to the case with lower
spin fields included. Note that a truncation of  lower spin fields
is only possible  at the cubic level. This is because at the cubic
level such an incomplete system remains formally consistent
because one can switch out interactions among any three elementary
(i.e., irreducible at the free field level) fields without
spoiling the consistency  at this order. This is a simple
consequence of the Noether current interpretation of the cubic
interactions: setting to zero some of the fields is always
consistent with the conservation of currents. However, lower spin
fields have necessarily to be introduced in the analysis of
higher-order corrections.

In what follows we formulate the action for the $AdS_5$ massless
boson and fermion gauge fields of $cu(1,1|8)$ that describes
properly higher-spin-gravi\-tatio\-nal interactions of spin $s\geq
3/2$ fields in the first nontrivial order. The result extends
the purely bosonic analysis (${\cal N}=0$) of \cite{VD5} to
the ${\cal N}=1$ supersymmetric case.

An appropriate ansatz for the higher spin action is of the form:
\be
\label{acta}
{\cal S}(R,R)=\frac{1}{2}{\cal A}(R,R)\,,
\ee
where the symmetric bilinear ${\cal A}(F,G)={\cal A}(G,F)$ is
defined for any 2-form higher spin curvatures (\ref{Rosn}) $F$ and $G$ in the adjoint of $cu(1,1|8)$
\be
\ba{l}
F=F_{E_1}(a,b)+F_{O_1}(a,b)\psi+F_{O_2}(a,b)\bpsi +F_{E_2}(a,b)\psi\bpsi\,,
\\
G=G_{E_1}(a,b)+G_{O_1}(a,b)\psi+G_{O_2}(a,b)\bpsi +G_{E_2}(a,b)
\psi\bpsi\,
\ea
\ee
as
\be
\label{act}
\dps {\cal A}(F,G)= {\cal
B}(F_{E},G_{E})+ {\cal F}(F_{O},G_{O})\;,
\ee
where \cite{VD5,Alkalaev:2002rq,A1}
\be
\label{bosact}
\dps
{\cal B}(F_{E},G_{E})\equiv{\cal B}^{\prime}(F_{E_1},G_{E_1})
+{\cal B}^{\prime\prime}(F_{E_2},G_{E_2})\;,
\ee
\be
\label{bosact2}
\ba{c}
\dps{\cal B}^{\prime}(F_{E_1},G_{E_1})=
 \int_{{\cal M}^5} \hat{H}_{E1}\wedge {\rm tr}
(F_{E_1}(a_1,b_1)\wedge G_{E_1}(a_2,b_2))|_{a_i=b_i=0}\,,
\\
\dps {\cal B}^{\prime\prime}(F_{E_2},G_{E_2})= \int_{{\cal M}^5}
\hat{H}_{E2}\wedge {\rm tr} (F_{E_2}(a_1,b_1)\wedge
G_{E_2}(a_2,b_2))|_{a_i=b_i=0}\,,
\ea
\ee
\be
\label{fermact}
\ba{r}
\dps {\cal F}(F_{O},G_{O})=\frac{1}{2}\int_{{\cal M}^5} \hat{H}_{O}\wedge
{\rm tr}(G_{O_2}(a_1,b_1)\wedge F_{O_1}(a_2,b_2))|_{a_i=b_i=0}
\\
\dps +\frac{1}{2}\int_{{\cal M}^5} \hat{H}_{O}\wedge {\rm
tr}(F_{O_2}(a_1,b_1)\wedge G_{O_1}(a_2,b_2))|_{a_i=b_i=0}\,.
\ea
\ee
1-forms $\hat{H}_{E1}, \hat{H}_{E2}, \hat{H}_{O}$ are the
following differential operators
\be
\label{He}
\ba{c}
\dps
\hat{H}_{E_i}= \alpha_i(p,q,t) E_{\alpha\beta} \frac{\d^2}{\d
a_{1\alpha} \d a_{2\beta}}\hat{b}_{12} +\beta_i(p,q,t)
E^{\alpha\beta} \frac{\d^2}{\d b_1^\alpha \d
b_2^\beta}\hat{a}_{12}
\\
+\dps \gamma_i(p,q,t)(E_\alpha{}^\beta \frac{\d^2}{\d a_{2\alpha}
\d b_1^\beta}\hat{c}_{21} -E^\alpha{}_\beta \frac{\d^2}{\d
b^\alpha_1 \d a_{2\beta}}\hat{c}_{12})\;,\quad i=1,2\,,
\ea
\ee
\be
\label{Ho}
\ba{c}
\dps \hat{H}_{O}=
\dps\alpha_3(p,q,t) E_{\alpha\beta} \frac{\d^2}{\d a_{1\alpha} \d
a_{2\beta}}\hat{b}_{12}\hat{c}_{12} +\beta_3(p,q,t)
E^{\alpha\beta} \frac{\d^2}{\d b_1^\alpha \d
b_2^\beta}\hat{a}_{12}\hat{c}_{12}
\\
\dps+\gamma_3(p,q,t) E_\alpha{}^\beta \frac{\d^2}{\d a_{1\alpha}
\d b_2^\beta}\;.
\ea
\ee
Here $E^{\alpha\beta}=DV^{\alpha\beta}$ is the frame field
defined by virtue of the compensator
$V^{\alpha\beta}(x)=-V^{\beta\alpha}(x): V^{\alpha\gamma}V_{\beta\gamma} =\delta^\alpha_{\beta}$ \cite{VD5}. The compensator is introduced
to define the Lorentz subgroup in $SU(2,2)$ in a manifest $AdS$ covariant manner.
The matrix $V^{\alpha\beta}$ is also used to
rise and lower spinor indices \cite{VD5}.
The coefficient functions $\alpha(p,q,t),\beta(p,q,t),\gamma(p,q,t)$,
which  parameterize various types of index contractions between two curvatures and the frame field
in (\ref{bosact2})-(\ref{fermact}), depend on
the operators:
\be
\label{t}
p=\hat{a}_{12}\hat{b}_{12}\;,
\qquad
q=\hat{c}_{12}\hat{c}_{21}\;,
\qquad t=\hat{c}_{11}\hat{c}_{22}\;,
\ee
where
\be
\label{abg}
\ba{ccc}
\dps
\hat{a}_{12} =
V_{\alpha\beta}\frac{\d^2}{\d a_{1\alpha} \d a_{2\beta}}\;,&
\qquad\dps \hat{b}_{12} = V^{\alpha\beta}\frac{\d^2}{\d b_1^\alpha
\d b_2^\beta}\;,& \qquad \dps\hat{c}_{ij} = \frac{\d^2}{\d
a_{i\alpha} \d b_j^\alpha}\;.
\ea
\ee
Potentially, in our analysis the higher spin gauge fields
are allowed to take values in some associative (e.g., matrix) algebra $\Omega
\rightarrow \Omega_I{}^J$. The resulting ambiguity is equivalent
to the ambiguity of a particular choice of the Yang-Mills gauge
algebra in the spin 1 sector. The classification of the higher
spin gauge theories associated with the different Yang-Mills
algebras is given in \cite{Vc}. Therefore, the higher spin action
(\ref{acta}) is
formulated in terms of the trace $tr$ in this matrix algebra (to
be not confused with the trace in the star product algebra).
Note that the gravitational field is required to take values in
the center of the matrix
algebra, being proportional to the unit matrix. For this reason,
the factors associated with the gravitational field are usually
written outside the trace.

For general coefficient functions $\alpha,\beta,\gamma$, the quadratic part of the action
(\ref{acta}) does not describe massless higher spin fields because
of ghost-type degrees of freedom associated with extra fields.
To eliminate these extra degrees of
freedom one should fix the operators $\hat{H}$ (\ref{He}) and
(\ref{Ho}) in a specific way by requiring the variation of the
quadratic action with respect to the extra fields to vanish
identically \cite{lv}. This condition is referred to as the
{\it extra field decoupling condition}. Another restriction on the
form of the action (\ref{acta}) comes from the requirement that
its quadratic part should decompose into an infinite sum of free
actions for different copies of fields of the same spin associated
with the spinor traces. This {\it factorization condition}
fixes a convenient basis in the space of fields rather
than imposes true dynamical limitations on form of the action
(in fact, this is the convenient diagonalization of the action).
Also, we introduce the {\it C-invariance condition} \cite{VD5,Alkalaev:2002rq} which
states that the action (\ref{act}) possesses the cyclic property
with respect to the central element of the higher spin superalgebra:
\be
\label{fcic}
{\cal A} (N\star F,G) = {\cal A}(F,G\star N)\;.
\ee
Being imposed, the condition simplifies greatly the analysis of
the dynamical system involving infinite sequences of supermultiplets of
the same spin. We show that the {\it extra field decoupling condition} along with
the {\it factorization condition} and the {\it C-invariance condition} fix
the coefficient functions $\alpha,\beta,\gamma$ up to arbitrary
functions parameterizing the ambiguity in the normalization
coefficients in front of the individual free bosonic and fermionic
actions \cite{Alkalaev:2002rq}.
{\it On the interacting level these functions are fixed unambiguously by the gauge invariance
 up to an overall factor in front of the action (\ref{acta})}.

The analysis of the gauge invariance in the cubic order is essentially based on
the statement called the First On-Mass-Shell Theorem
\cite{FV1,VD5,Alkalaev:2002rq}. It states that most of linearized
higher spin curvatures components are zero on mass-shell except for
those corresponding to Weyl tensors $C$, {\it i.e.} (schematically)
\be
\label{R=C}
R_1=h\wedge h C + X(\frac{\delta {\cal S}_2}{\delta\Omega})\,,
\ee
where $h$ stands for the background value of the frame field $E$ and $X$ are some linear functionals
of the free equations of motion.
The condition that the {\it cubic} higher spin
action is invariant under some deformation of the higher spin
gauge transformations is {\it equivalent} to the condition that
the original (i.e. undeformed) higher spin gauge variation of the
action is zero once the linearized higher spin curvatures $R_1$
are replaced by the Weyl tensors $C$. As a result, the problem is
to find such functions $\alpha, \beta$ and $\gamma$ that
\be
\label{SEtr}
\delta {\cal S} (R,R)  \Big
|_{E=h, R = h\wedge h C} \equiv  {\cal A}^h_{\alpha,\beta,\gamma}
(R,[R,\epsilon ]_\star)
 \Big |_{R = h\wedge h C} =0\,
\ee
for an arbitrary gauge parameter $\epsilon(a,b,\psi,\bpsi|x)$.
As to the terms in the variation which involve $X$-dependent parts of
higher spin curvatures (\ref{R=C}), these are exactly compensated
by appropriate deformations of the abelian higher spin gauge transformations (\ref{gotr}).
These deformation corrections do not vanish on mass-shell.
The non-trivial part of the variation is therefore reduced to the condition (\ref{SEtr}).
It is crucial to stress that consideration of higher orders in interactions
essentially requires curvature-dependent corrections (yet unknown) in
(\ref{R=C}) which can be disregarded in the cubic order.

Note that terms resulting from the gauge transformations of the
gravitational fields and the compensator $V^{\alpha\beta}$
contribute into the factors in front of the higher spin curvatures
in the action $(\ref{act})-(\ref{fermact})$. The proof of the
respective invariances is given in \cite{VD5} and is based
entirely on the explicit $su(2,2)$ covariance and invariance of
the whole framework under diffeomorphisms. Also, one has to take
into account that the higher spin gauge transformation of the
gravitational fields is at least linear in the dynamical fields
and therefore has to be discarded in the analysis of $\Omega^2
\epsilon $ type terms under consideration.

As shown in \cite{Alkalaev:2002rq}, the condition (\ref{SEtr})
along with  the {\it extra field decoupling
condition}, the {\it factorization condition} and the {\it C-invariance condition} fix the
coefficient functions in the form
\be
\label{gab11}
\alpha_1(p,q,t) +
\beta_1(p,q,t)  =  \Phi_0  \sum_{m,n =0}^\infty
\frac{(-1)^{m+n}(m+1)\; p^n q^m}{2^{2(m+n+1)} (m+n+2)! m! (n+1)!}\,,
\ee
\be
\label{gab12}
\gamma_1(p,q,t)  = \gamma_1(p+q) \,,\quad
\gamma_1(p)=\Phi_0  \sum_{m =0}^\infty
\frac{(-1)^{m+1}\;p^m}{2^{2m+3} (m+2)! m!} \,,
\ee
\be \label{gab21} \alpha_2(p,q,t) + \beta_2(p,q,t)  =
\frac{1}{4}(\alpha_1(p,q,t) + \beta_1(p,q,t))\,, \quad
\gamma_2(p,q,t)  = \frac{1}{4}\gamma_1(p,q,t)\,,
\ee
\be
\label{gab31}
\alpha_3(p,q,t)+\beta_3(p,q,t)=
\Phi_0\sum_{m,n=0}^{\infty}\frac{(-1)^{m+n+1}\;p^m\,q^n}{2^{2(m+n)+3}\,(m+1)!\,(m+n+2)!\,n!}\;,
\ee
\be
\label{gab32}
\gamma_3(p,q,t)=\gamma_3(p+q)\;,\quad\gamma_3(p)=
\Phi_0\sum_{m=0}^{\infty}(-1)^{m+1}\frac{(-1)^{m+1}\;p^m}{2^{2m+1}\,m!\,(m+1)!}\;,
\ee
where $\Phi_0 $ is an arbitrary normalization factor to be
identified with the (appropriately normalized in terms of the
cosmological constant) gravitational coupling constant.

The consistent cubic interactions can also be constructed in the framework
of the model exhibiting higher spin symmetries based on
the algebra $hu_0 (1,1|8)$ that results  from  factoring out
the maximal ideal in $cu(1,1|8)$ \cite{Vc}.
Elements of this algebra are spanned by the supertraceless multispinors.
Thus the $hu_0(1,1|8)$ describes the system of higher spin fields with every
supermultiplet emerging once. One can consider further reduction of $cu(1,1|8)$  and
introduce the higher spin algebra $ho_0(1,1|8)$ along with its bosonic subalgebra
$ho_0(1,0|8)$ studied in \cite{Vc,ss}.

The consideration of the reduced model is based on
the approach elaborated for the pure bosonic system in
\cite{VD5} which consists of inserting a sort of projection
operator ${\cal M}$ to the quotient algebra into the action (\ref{acta}).
Namely, let ${\cal M}$ satisfy $N\star {\cal M} ={\cal M}\star N=0$.
Having specified the "operator"
${\cal M}$ we write the action for the reduced system associated
with $hu_0 (1,1|8)$ by replacing the bilinear form in the action
with
\be
\label{bformc}
{\cal A}(F,G) \to {\cal A}_0 (F,G)=  {\cal A}(F,{\cal M}\star G)\,,
\ee
where ${\cal A}(F,G)$ corresponds to
the action  describing the original (unreduced) higher spin
dynamics. {}From the {\it C-invariance condition} it follows
$ {\cal A}(F, {\cal M}\star G)={\cal A}(F\star
{\cal M}, G)$
provided that for any $F,G \in cu(1,1|8)$ the operator ${\cal M}$ satisfies
$
F\star {\cal M}= {\cal M}\star F,\,G\star {\cal M}={\cal M}\star G\,.
$
In fact, this implies that the operator ${\cal M}$ should be a function of $N$.
As a result, all terms proportional to $N$ do
not contribute to the action (\ref{bformc}) which therefore is
defined on the quotient subalgebra. The representatives of the
quotient  algebra $hu_0(1,1|8)$ are identified with the elements
$F$ satisfying the supertracelessness condition
$
P^-F=0.
$
We find the operator ${\cal M}$ in the form
\be
\label{FINAL}
{\cal M}=\sum_{n=0}^{\infty}\,\frac{N^{ 2n}}{n!\,(n+1)!}\;.
\ee
The modification of the action according to (\ref{bformc}) does
not contradict to the previous conclusions, where the
action (\ref{acta}) was claimed to be fixed unambiguously, because
in that analysis we have imposed the {\it factorization condition}
in the particular basis of higher spin fields thus not allowing
the transition to the invariant action (\ref{bformc}). The {\it
factorization condition} is relaxed here. All other
conditions, namely the {\it C-invariance condition}, the {\it extra field
decoupling condition} and the condition (\ref{SEtr}) remain valid.

It is not straightforward to incorporate an extended supersymmetry with
$\cal{N}\geq $ $2$  in the present construction of cubic
higher spin couplings. This is because $\cal{N}\geq $ $2$ supermultiplets
originated from $cu(2^{{\cal N}-1}, 2^{{\cal N}-1}|8 )$ require mixed-symmetry
higher spin fields to be included \cite{VD5,Vc}.
Two comments are in order. Firstly,
the method employed in the present paper for
constructing higher spin cubic couplings is essentially based on
Lagrangian formulation of  symmetric higher spin gauge fields in the form
discovered in \cite{lv}.
Secondly, such Lagrangian description of mixed-symmetry
massless gauge fields in $AdS_d$ with $d\geq 5$ is not elaborated yet.
Let us note, that advantages of the approach of \cite{lv}
are manifest higher gauge invariance of quadratic actions and the
First On-Mass-Shell theorem which, on the cubic level, effectively
determines true dynamical components of higher spin curvatures (Weyl tensors)
thus solving the problem of ghost-type degrees of freedom.
Moreover, the approach has  manifest invariance under space-time
diffeomorphisms.

To conclude, let us note that although we have proved
the existence of particular higher-spin-gravitational couplings in
the first order in interaction,
there are many vital problems which remain unsolved. For instance, one of interesting question
for further investigations is the study of the cubic couplings of
particular higher spins in the form suitable for applications.
Being rather non-trivial, this problem is important in context of
higher spin $AdS/CFT$ correspondence.
However, in the author's humble opinion, only the knowledge of the higher
spin action in the non-perturbative form will bring to light
an adequate language to analyze the holography and the relationship with string theory
and all the related topics.

\vspace{5mm}

\noindent \textbf{Acknowledgments.}
The material presented here is based upon work done in collaboration
with M.A. Vasiliev.
This research was supported in part by RFBR Grants No 02-02-17067,
No 02-02-16944 and the Landau Scholarship Foundation,
Forschungszentrum J\"u\-lich.

\end{document}